\documentstyle[12pt]{article}
\title{Particle physics from the noncommutative geometry point of view
$\dag$ *}
\author{J. S\l adkowski \\
\sl Institute of Physics, University of Silesia,\\
\sl 40-007 Katowice ul. Uniwersytecka 4, Poland, \\
and \\
\sl Fakult\"at f\"ur Physik, Universit\"at Bielefeld,\\
\sl P.O. Box 100131, D-33501 Bielefeld, Germany.}
\begin{document}
\date{}
\baselineskip20pt
\maketitle
\begin{abstract}
\ \ \ Recent development in noncommutative geometry
generalization of gauge theory is reviewed. The mathematical
apparatus is reduced to minimum in order to allow the
non-mathematically oriented physicists to follow the development
in the interesting field of research.
\end{abstract}

\vspace{50mm}
\ \ \ $\dag$ Partially supported by Deutsche Forschungsgemeinschaft. \\

\ \ \ * Invited talk presented at the Silesian School of Theoretical
Physics: Standard Model and Beyond, Szczyrk (Poland), September 1993.

\newpage
\section{Introduction}

\ \ \ The unification of electromagenetic and weak interactions is one of
the biggest achievements of theoretical physics (the GWS model). This model
successfully describes all known experiments involving electroweak
interactions, although the gauge sector is not yet directly accessible in
experiment [1]. We believe that the existence of the Higgs particle and the
missing members of the third family will be soon confirmed. The situation
is far less satisfactory  from the theoretical point of view because the
GWS model contains too many free parameters and the symmetry breaking
mechanism is not yet understood. Much research have been made into the
structure of string theories hoping to find answers to this questions [2].
Recently, new ideas that make use of the A. Connes' noncommutative geometry
have been put forward [3]. A. Connes managed to reformulate the standard
objects
of differential geometry in a pure algebraic way. This allows allows for
generalization of differential geometry to the more exotic cases of sets than
manifolds.  This new formalism has been immediately applied in gauge theory
because it allows for generalization of the Kaluza-Klein program to the
discrete internal space case.

\section{Main ideas of noncommutative geometry}

\ \ \  Mathematicians have proved that a given topological space X can be
equivalently described by the (commutative) algebra C(X) of real (complex
in the complex case) valued continuous
function on X. It is also possible to describe the standard notions of
differential geometry in terms of algebraic structures on C(X). We have the
following correspondence (as this review is aimed at
non-mathematically-oriented
readers we will not give the precise definitions that we will not need; they
can
be found in [3, 4]):
$$\begin{array}{lcl}
topological \ space\ X &\equiv & C(X) \cr
manifold \ M &\equiv &  C^{\infty} (M) \cr
vector\ bundle\ on\ M& \equiv & projective\ module\ over \ C^{\infty}(M)\cr
connection &\equiv& "universal" \ connection\cr
\end{array} $$
The positive answer to the question {\it can one go further and get rid of the
adjective commutative in front of the algebra in question?} was given by A.
Connes [3]. The result of this generalization, referred to as noncommutative
geometry, allows us to do differential geometry more sophisticated level. As
differential geometry is widely used in theoretical physics, it is not
surprising that the newly invented noncommutative geometry became a very
promising tool in physicists' hands. Here we will restrict ourselves to the
particle physics. To "do the noncommutative particle physics", one have to
specify the fermionic content of the theory and the gauge group. One
introduces fermions by defining an appropriate Dirac operator. The gauge group
can be a priori arbitrary but for technical reasons only unitary groups of
the algebra $A$ that generalizes $C(X)$

$$ U_{n}=\{ u \varepsilon M_{n}\left( A\right) : uu^{\dagger}=u^{\dagger}
u=1 \} \ ,\eqno(1)$$
where $M_{n}(A)$ is the $n\times n$ matrix with entries from $A$
fit naturally to the
formalism. The gauge group is defined by giving "an extension" of the algebra
of function on the (approximate?) spacetime. To be more precise, let us
define:
$$ Definition \ 1.$$

Given an arbitrary algebra $A$, we can construct an algebra $\Omega
A$ as follows. To every element $a$ $\varepsilon $ $A$ we associate  a new
element $da$. As a vector space, $\Omega A$ is the linear space of words built
out of the "letters" $a$ and $da$. Multiplication of two such words is
performed
by concatenation and one imposes the associativity and distributivity over
the action "+". Further, we will require that
$$ d1=0\ , \ d\left( a_{0}a_{1} \right) - da_{0}a_{1} - a_{0}da_{1}=0\ ,\
 and \ d^{2}=0.\eqno(2)$$
This is a very abstract notion. To make it more mundane, let us represent it
in a (physical) Hilbert space H by setting (we neglect the very mathematical
subtleties such as existence, correctness and so on) via

$$ \pi \left( a_{0}da_{1}\dots da_{n}\right) \equiv i^{n}a_{0}
\left[ D,a_{1}\right] \dots \left[ D,a_{n}\right]\ ,\eqno(3)$$
where $D$ is the free Dirac operator.  In the physically motivated cases,
H is a $Z_{2}$-graded space, equipped with a grading operator $\Gamma$
($\gamma _{5}$-matrix) such that $\Gamma ^{2}=1$, $A$ acts on H by even
operators
and D is an odd operator, ie:

$$ a\Gamma=\Gamma a\  for\  a \varepsilon A \ and \ D\Gamma = -\Gamma D$$
Bellow, we will ignore the precise structure of the spacetime and focus our
attention on the appropriate algebraic structures. This simplify our task,
although, we will loose the geometrical interpretation.

$$ Definition \ 2.$$

A gauge field (connection) is any (skew) form $\alpha \varepsilon
\Omega ^{1}A $, $ \alpha = \sum a^{i}db^{i}$ such that $\sum a^{i}b^{i} =1$.
It determines the covariant derivative $\nabla = d + \alpha$.
The curvature (stress tensor) is given by $\Theta =d\alpha + \alpha ^{2}$.

Now, we have [4, 5, 6]

$$ {\it L}_{YM}= \frac{1}{8} \int _{M} Tr\left( \pi \left( \Theta \right)
^2\right)\eqno(4)$$

$${\it L}_{F}=\int _{M} <\psi| D+ \pi \left( \alpha \right) |\phi >\ ,
\eqno(5)$$
where $ {\it L}_{YM}$ and ${\it L}_{F}$ denote the bosonic and fermionic
parts of the Lagrangian, respectively.

\section{Models}

\ \ \ Let $S$ be a Riemannian (spin) 4-manifold, $N_{G}$ denote the number
of generations, $M_{IJ}$ be the $N_{G}\times N_{G}$  ($I,J= 1,2,3,4$)
"mass matrices" and

$$D=\pmatrix{\not \partial \otimes Id & \gamma _{5} \otimes M_{12}& \gamma _{5}
\otimes M_{13} & \gamma _{5} \otimes M_{14}\cr \gamma _{5} \otimes M_{21}
&\not \partial \otimes Id & \gamma _{5} \otimes M_{23}& \gamma _{5} \otimes
M_{24}\cr
\gamma _{5} \otimes M_{31} & \gamma _{5} \otimes M_{32}& \not \partial \otimes
Id
& \gamma _{5} \otimes M_{34}\cr  \gamma _{5} \otimes M_{41}& \gamma _{5}
\otimes M_{42}& \gamma _{5} \otimes M_{43}&\not \partial \otimes Id \cr}.
\eqno(6)$$
Here, the matrices $M_{IJ}$ describe the fermionic mass sector including
mixing [7, 8]. Let $A= C(S)\otimes {\bar A}$, where ${\bar A}$ is the algebra

$${\bar A}= M_{n_{1}}\oplus M_{n_{2}}\oplus M_{n_{3}}\oplus M_{n_{4}}\eqno(7)$$
of direct sum of complex $n_{i}\times n_{i}$ matrices. An element $a
\varepsilon A$ can be written as

$$a= diag \left( a_{1}, a_{2}, a_{3}, a_{4}\right) \ , \eqno(8)$$
where $a_{i} \varepsilon M_{n_{i}}\left( C \left( S\right) \right)$, where
the matrices are "built out" of complex function on spacetime.
We have to compute the gauge field

$$\pi \left( \alpha \right) = \sum _{i} a^{i}\left[ D,b^{i}\right] \
.\eqno(9)$$
Simple calculation leads to

$$\left[ D,b^{i}\right] = \pmatrix{\not \partial b^{i}_{1}& \gamma _{5}\otimes
\left( M_{12}b^{i}_{2}-b^{i}_{1}M_{12}\right) &\gamma _{5}\otimes
\left( M_{13}b^{i}_{3}-b^{i}_{1}M_{13}\right) & \dots \cr
\gamma _{5}\otimes
\left( M_{21}b^{i}_{1}-b^{i}_{2}M_{21}\right) & \gamma _{5}\otimes \not
\partial b^{i}_{2}
 & \dots &\dots \cr
\vdots & \vdots & \vdots &\vdots \cr \gamma _{5}\otimes
\left( M_{41}b^{i}_{1}-b^{i}_{4}M_{41}\right) & \dots &\dots & \not
\partial b^{i}_{4} \cr}\eqno(10)$$
So that

$$\pi \left( \alpha \right) = \pmatrix{ A_{1} & \gamma _{5} \otimes
\phi _{12} & \gamma _{5} \otimes
\phi _{13} & \gamma _{5} \otimes
\phi _{14} &\cr\gamma _{5} \otimes
\phi _{21} & A_{2} & \gamma _{5} \otimes
\phi _{23} &\gamma _{5} \otimes
\phi _{24} & \cr \gamma _{5} \otimes
\phi _{31} &\gamma _{5} \otimes
\phi _{32} & A_{3}&\gamma _{5} \otimes
\phi _{34} & \cr \gamma _{5} \otimes
\phi _{41} &\gamma _{5} \otimes
\phi _{42} &\gamma _{5} \otimes
\phi _{43} & A_{4} \cr}\ , \eqno(11)$$
where

$$A_{m}= \sum_{i}a^{i}_{m} \not \partial b^{i}_{m} \eqno(12)$$
and

$$\phi_{mn}= \sum_{i} a^{i}_{m} \left( M_{mn}b^{i}_{n} -b^{i}_{m}M_{mn}
\right) \ .\eqno(13)$$
Further, we have to calculate

$$\pi \left( d\alpha \right) =\left[ D, \alpha \right] =\sum _{i}
\left[ D a^{i}\right]\left[ D,b^{i}\right] \eqno(14)$$
and

$$\pi \left( \Theta \right)=\pi \left( d\alpha + \alpha ^{2}\right)
\ .\eqno(15) $$
This leads to

$$\pi \left( \Theta \right)_{mm}= \frac{1}{2} \gamma^{\mu \nu}F_{\mu \nu}
^{m} + \sum_{p\neq m}|K_{mp}|^{2}|\phi_{mp}+ M_{mp}|^{2} -Y_{m} -X'
_{mm}\eqno(16)$$
where

$$X'_{mm}= \sum_{i} a^{i}_{m}{\not \partial} ^{2} b^{i}_{m}\eqno(17)$$

$$Y_{m}=\sum_{p\neq m}\sum_{i}a^{i}_{m}|K_{mp}|^{2}|M_{mp}|^{2}
b^{i}_{m}\eqno(18)$$

$$F^{m}_{\mu \nu}=\partial _{\mu}A_{\nu}^{m}-\partial _{\nu}A_{\mu}^{m}+
\left[ A_{\mu}^{m} , A_{\nu}^{m}\right] \ . \eqno(19)$$
Here, we have "generalized", following [8], the matrices $M_{ij}$

$$ M_{ij} \rightarrow K_{ij}\otimes M_{ij} \ . \eqno(20)$$
Now, $K_{ij}$ describes the mixing among families and $M_{ij}$ describes the
vacuum expectation values of the Higgs sector. The off-diagonal elements are
given by

$$\begin{array}{lll}
\pi \left( \Theta \right)_{mn}& = &-\gamma _{5} K_{mn}\left( {\not \partial}
\phi _{mn} + A_{m} \left( \phi _{mn} +M_{mn}\right) -\left( \phi _{mn} +M_{mn}
\right) A_{n}\right) -X_{mn}\cr  & & \mbox{} + \sum_{p\neq m,n} K_{mp}K_{pn}
\left( \left(

\phi _{mp} + M_{mp}\right) \left( \phi _{pn} + M_{pn} \right) -
M_{mp}M_{pn} \right) \ ,
\end{array} \eqno(21)
$$
where

$$X_{mn}=\sum _{i}\sum _{p\neq m,n}K_{mp}K_{pn}\left( M_{mp}M_{pn}b^{i}_{n}
- b^{i}_{m}M_{mp}M_{pn}\right) \ .$$

This completes the bosonic sector of the model:

$$\begin{array}{lll}
{\it L_{YM}}&=&-\sum _{m=1}^{m=4}Tr \left( \frac{1}{4} F_{\mu \nu}^{m}
F^{m\ \mu \nu}-\frac{1}{2} |\sum _{p\neq m}\left( |K_{mp}|^{2}|\phi _{mp}
+ M_{mp}|^{2} -Y_{m} \right) - X'_{mp}|^{2} \right.\cr & & \mbox{} +
\frac{1}{2}
\sum _{p\neq m}|K_{mp}|^{2}
|\partial _{\mu} \left(  \phi _{mp}+ M_{mp}\right) + A_{\mu}^{m}\left( \phi
_{mp} +
M_{mp} \right) -\left( \phi _{mp} +M_{mp} \right) A_{\mu}^{p} |^{2}  \cr
& & \left. \mbox{} -\frac{1}{2}\sum _{n\neq m}\sum _{p\neq m,n} | |K_{mp}|
^{2}\left( \left( \phi _{mp} +
M_{mp} \right) \left( \phi _{pn} + M_{pn} \right) - M_{mp}M_{pn}\right) -
X_{mn}|^{2} \right)  \ .
\end{array} \eqno(22)$$

\section*{{\bf 3A.}\ The $SU(2)\times U(1)$ model}

\ \ \ To obtain the $SU(2)\times U(1)$ electroweak unification model we should
consider the algebra of the form $ M_{2\times 2}\oplus M_{1\times 1}$,
where $M_{i\times i}$ are the $i \times i$ matrices over the ring
of complex valued function on spacetime. This aim can be also
achieved by considering the extension of spacetime of the form $S
\times \{1,2\}$, the product of the ordinary spacetime $S$ by a
two-point set [5, 6].

We choose the following free  Dirac operator [8]
$$D=\pmatrix{{\not \partial}\otimes 1&\gamma_{5}\otimes M_{12}\cr
\gamma_{5}\otimes M_{21}&{\not \partial}\otimes 1\cr}\eqno(23)$$
with the mass matrix of the form

$$M_{12}= \pmatrix{0\cr \mu} =S\eqno(24)$$
By repeating the above calculation we get

$$\pi \left( \alpha \right)=\pmatrix{A_{1I}^{J} & H^{J}\cr
H_{I} & A_{2}\cr}\eqno(25)$$
$$H=\mu \sum _{i} a_{1}^{i}\left( Sb^{i}_{2} - b^{i}_{2}S\right) \eqno(26)$$

$$a_{1} \varepsilon M_{2\times 2}\left( C^{\infty}\left( M\right) \right) \ , \
\
a_{2} \varepsilon M_{1\times 1}=
C^{\infty}\left( M\right) \eqno(27)$$
In addition, we will demand that

$$Tr A_{1}=Tr A_{2}=A_{2}\eqno(28)$$
in order to reduce the gauge group from $U(2)\times U(1)  $ to $SU(2)
\times U(1)$.

The auxiliary fields take the form

$$X_{12}=X_{21}=0 \ ,\ \ Y_{2}= \mu ^{2} \eqno(29)$$
$$Y_{1} = \mu ^{2} \sum _{i} a^{i}_{1} Tb^{i}_{1}\eqno(30)$$
$$ T=\pmatrix{0&0\cr 0&1}\eqno(31)$$
and can be easily get rid off. Finally, we get [8]

$$\begin{array}{lll}
{\it I_{YM}}&=& \frac{1}{4}\left( \left( F_{\mu
\nu}^{1}\right) ^{I}_{J} \left( F^{1\mu \nu}\right) ^{J}_{I} + F_{\mu
\nu}^{2}F^{2\mu \nu} \right) \cr & & + \frac{1}{2} Tr KK^{\dag}| \partial
_{\mu} \left( H^{I} + H_{0}^{I}\right) + A^{1I}_{\mu J}\left( H^{J}
+H_{0}^{J} \right) - \left( H^{I} + H_{0}^{I}\right) A_{\mu}^{2}|^{2}
\cr & & - \frac{1}{2}\left( Tr \left( KK^{\dag}\right) ^{2} -\left(
TrKK^{\dag} \right) ^{2}\right) \left(  \left( H^{I} +
H_{0}^{I}\right) \left( H^{\dag}_{I} + H^{\dag}_{0I}\right) - \mu
^{2}\right) ^{2}
\end{array}\ . \eqno(32)$$
The fermionic sector has the form

$$\begin{array}{lll}
{\it L_{f} }&=& {\bar l}_{L}\left( D + \pi \left( A\right) \right)
l_{L} + {\bar e}_{R}\left( {\not \partial} + A_{2}\right) e_{R}\cr
& & +{\bar l}_{L} \left( H +H_{0}\right) e_{R}K + {\bar e}_{R}\left(
H^{\dag} +H^{\dag}_{0}\right) l_{L}K^{\dag}
\end{array} \eqno(33) $$

To get a realistic model we have to include the strong interaction.
"Unfortunately" the colour gauge group is unbroken. This makes the
things more complicated because unbroken gauge symmetries are not in
the spirit of the noncommutative geometry approach. To this end, we
have to extend the gauge group by the $SU(3)\times U(1)$ factor and
identify the two $U(1)$ factors (sort of charge quantization
condition can be deduced from this [4]).

\ \ \ The left-right symmetric model can be also constructed [8,
9].
The complications connected with the $SU(3)_{colour}$ factor suggest
that grand unified models are more natural then the "partial
unification" in the noncommutative framework.

\section*{{\bf 3B.}\ Grand unification}

\ \ \ The discussed here formalism can be easily applied to grand
unification. If one considers the algebra

$$ C^{\infty} \left( S\right) \otimes \left( M_{5\times 5} \left( \bf C\right)
\oplus M_{5\times 5}\left( \bf C\right) \oplus M_{1\times
1}\left( \bf R\right) \right) \ ,\eqno(34)$$
where ${\bf C}$ and ${\bf R}$ denote the complex and real
numbers, and demands the permutation symmetry between the two $
M_{5\times 5}$ terms, one gets

$$
\pi \left( \rho \right)= \pmatrix{A & \Sigma & H\cr \Sigma & A
&H\cr H^{*} & H^{*} & 0} \ .\eqno(35)$$
Here, $ H$ is a complex scalar field  and $\Sigma $ a $5\times
5$  self adjoint scalar field. One have to force the condition
$Tr A= 0$ on the gauge field $A$ in order to reduce the gauge group
from $U(5)$ to $SU(5)$. One can find such values for the mass
matrices

$$ M_{12}=M_{21}=\Sigma _{0}\eqno(36)$$
and

$$ M_{13}=M_{23}=H_{0}\eqno(37)$$
so that interesting, although, phenomenologically unacceptable
$SU(5)$ GUT models are "produced" [8]. The more natural choice
of $M_{1\times 1}\left( \bf C\right) ={\bf C} $ instead of $M_{1\times
1}\left( \bf R\right) ={\bf R}$ in (34) leads to noncommutative analogues
of the "flipped unification" models. Such models might result in
a phenomenologically acceptable model. In the seminal paper [10],
it was shown that the $SO(10)$ GUT is also possible in the
noncommutative framework! One have to consider the algebra

$$P_{+} Cliff \left( SO \left( 10 \right) \right) P_{+}\oplus {\bf
R} \ , \eqno(38)$$
where

$$P_{+}=\frac{1}{2}\left( 1+\Gamma _{11}\right) \eqno(39)$$
as the factor that extent the algebra of function on spacetime.

\section*{{\bf 3C.}\ Nonlinear Higgs mechanism}

\ \ \ Here we would like to point out that the noncommutative
generalization of gauge theory may predict a nonlinearly
realized spontaneous
symmetry breaking, known under the acronym BESS (breaking
electroweak sector strongly) [11-13]. Our main argument for
BESS can be stated as follows. The noncommutative version of
the standard model predicts the required form of the Higgs
sector but fermion masses (Yukawa couplings) and the number
of generation, $N_{G}$, are free parameters. There must be at
least two generations but why not, say, 127? It is natural to
suppose that $N_{G}$ is big or even unlimited and that the
fermion masses emerge as a result of interaction and the spacetime
structure. We see only the lightest fermions because the energy at
our disposal is not high enough. The Higgs particle has not yet
been discovered. Does it really exist as a physical particle? We
will show that it can be thought of in the
limit $m_{H} \rightarrow
\infty$. The main argument against BESS is that such models are
nonrenormalizable. Noncommutative geometry says that our notion
of spacetime is only an approximation (an effective electromagnetic
spacetime). The correct description is in terms of algebras.
Should we not give up the requirement of renormalizability?
BESS models can certainly lead to physical prediction [14].
General relativity provide us with analogous arguments.
Following the rules described above, we can construct the
the Lagrangian of the Standard Model [6, 15]

$$\begin{array}{ll}
{\it L_{YM}}= &\int \lbrace \frac{1}{4} N_{G}\left(
F_{\mu \nu}^{1}F^{1 \mu \nu}  + F_{\mu \nu}^{2}F^{2 \mu \nu} +
F_{\mu \nu}^{c}F^{c \mu \nu} \right) \cr
\ & + \frac{1}{2}Tr\left( MM^{\dag}\right) | \partial
H + A_{1}H - H^{\dag}A_{2} | ^{2}\cr
 \  & -\frac{1}{2} \left( Tr\left( MM^{\dag}\right) ^{2} - \left(
Tr MM^{\dag}\right) ^{2}\right) \left( HH^{\dag} -1\right)
^{2}\rbrace d^{4}x \cr \ .
\end{array}\eqno(40)$$
The fermionic action is given by

$$
\begin{array}{lll}
{\it L_{f}} & = & <\psi | D + \pi \left( \rho \right) | \psi >\cr
\ & = & \int \left( {\bar \psi } _{L} {\bar D}\psi _{L} +
{\bar \psi } _{R} {\bar D}\psi _{R} + {\bar \psi }_{L}H
\psi _{R} + {\bar \psi }_{R}H^{\dag}
\psi _{L}\right) d^{4}x \end{array}\ , \eqno(41)$$
where we have included the diagonal part of $\pi (\rho)$ term into ${\bar D}$.

\ \ \  Let us look closer at the full Lagrangian, ${\it L = L_{YM}
+ L_{f}}$. It has the standard form except for the $N_{G}$ factor
in front of the gauge field kinetic terms that comes from the
trace over generations. The analogous term in ${\it L_{f}}$ give
the sum over generations. We know that there are only three light
generations of fermions but is that all? We should count all
generations in ${\it L}$! This means that the coefficient in front
of the $F_{\mu \nu}F^{\mu \nu}$ terms should depend on $N_{G}$
and, in fact, give us information about the total numbers of
generations because it is absent from the fermionic part!
This is not true. The orthodox normalization is correct. We should
normalize the Diximier trace [3, 4] that leads to (4, 5) so that the
coefficient
$N_{G}$ disappears. The simplest and natural solution is to
normalize $Tr$ so that $Tr Id_{N_{G}}=1$ [8]. This ensures also
that $Tr_{\omega}$ is always finite. There is a natural inner
product on the algebra of complex square matrices given by
$Tr(AB^{\dag})$. If one apply the Cauchy-Schwarz Inequality
to this inner product, one gets

$$Tr\left( MM^{\dag}\right) ^{2} \le \left( TrMM^{\dag}\right) ^{2}
\eqno(42)$$
We cannot ensure the correct sign of the Higgs mass term without
the above normalization. The normalization of the trace $Tr$ leads
to

$$Tr\left( MM^{\dag}\right)^{2} \le N_{G} \left( TrMM^{\dag}
\right)^{2} \ . \eqno(43)$$
This means that for a big $N_{G}$ the coefficient $K=Tr\left(
MM^{\dag}\right) ^{2} - \left( TrMM^{\dag}\right) ^{2}$ may be
very large. In fact, it is possible that $K \rightarrow \infty$ if
the number of heavy  generations is unlimited. This force the
condition $HH^{\dag}=1$ in the Lagrangian and removes the Higgs
particle from the spectrum! If we are going to interpret the
Yukawa coupling in the standard way then we are not allowed to
arbitrary rescale the Higgs field and the limiting case leads to

$$m_{H}=\sqrt {2{{Tr\left( MM^{\dag}\right) ^{2} -
\left( TrMM^{\dag}\right) ^{2}} \over {Tr\left( MM^{\dag}\right)
 }}}\  \
\rightarrow \infty \eqno(44)$$
as should be expected. The fermionic masses are generated in such a
(nonlinear) model by means of Yukawa couplings in a way analogous
to that of the standard model [11-13]. The fermionic part of the
Lagrangian given by Eq. (41) has the required form!

\section{Final remarks}

\ \ \ We have reviewed recent development in the noncommutative
particle physics. As we wanted to reduce the mathematical
apparatus to the minimum to make it accessible
non-mathematically oriented physicists,  we have neglected the
mathematical subtleties and the spacetime structure. The
interested reader is referred to [3, 4].\\

\ \ \ The complete understanding of the noncommutative particle
physics is impossible without quantization. Up to know, we are
able to get more or less interesting classical Lagrangian that
can be quantized in the usual way. But it may not be the correct
way of doing noncommutative physics! Toy model considerations
suggest that certain relations among physical variables
predicted by the classical Lagrangians are spoiled by quantum
correction. To get the Lagrangian, we have to get rid of the
"auxiliary fields" using equations of motion. Is it possible in
a quantum theory? If not, we should consider the the possibility
of condensation the bosonic sector along the lines considered in
[16]. In general we should expect relation among vev's of the
scalar and vector fields because in the noncommutative framework
thy are related. \\

\ \ \ There is also the question of possible extra terms that
are not allowed or vanish in the orthodox approach [17]. Such terms, if
found, may result in unexpected physical consequences. \\

\ \ \ {\bf Aknowledgement:} I greatly enjoyed the hospitality extended
to me during a stay at the Faculty for Physics at the University of
Bielefeld, where the final version of the talk was discussed and written
down. \\

\newpage

\section*{References}

\newcounter{bban}

\begin{list}
{[\ \arabic{bban}\ ]}{\usecounter{bban}\setlength{\rightmargin}
{\leftmargin}}

\item D. Schildknecht, in these proceedings.
\item J. S\l adkowski, Fortschr. Physik {\bf 38} (1990) 477.
\item A. Connes, Publ. Math. IHES {\bf 62} (1983) 44.
\item J. G. V\'arilly and J. M. Garcia-Bond\'ia, to be published in
J. Geom. Phys.
\item A. Connes, in The interface of mathematics and physics
(Claredon, Oxford, 1990) eds . D. Quillen, G. Segal and S. Tsou.
\item A. Connes and J. Lott, Nucl. Phys. {\bf B} Proc. Suppl.
{\bf 18B} (1990) 29.
\item A. H. Chamseddine, G. Felder and J. Fr\"ohlich, Phys. Lett.
{\bf B296} (1992) 109.
\item A. H. Chamseddine, G. Felder and J. Fr\"ohlich, Nucl. Phys.
{\bf B395} (1993) 672.
\item B. E. Hanlon and G. C. Joshi, Univ. of Melbourne preprint,
UM-P-92/38.
\item A. H. Chamseddine and J. Fr\"ohlich, ETH preprint,
ETH/TH/93-12.
\item R. Cosalbuoni, S. de Curtis, D. Dominici and R. Gatto,
Nucl. Phys. {\bf B282}, 235 (1987); Phys. Lett. {\bf B155},
(1985) 95.
\item G. Cvetic and R. K\"ogerler, Nucl. Phys. {\bf B328}
 (1989) 342.
\item G. Cvetic and R. K\"ogerler, Nucl. Phys. {\bf B353} (1991)
462.
\item R. B\"onish and R. K\"ogerler, Int. J. Mod. Phys.
{\bf A7} (1992) 5475.
\item J. S\l adkowski, Bielefeld University preprint,
BI-TP93/26 (1993).
\item R. Ma\'nka and J. Syska, to appear in Phys. Rev. D (1993).
\item A. Sitarz, Phys. Lett. {\bf B}308 (1993) 311.

\end{list}

\end{document}